\documentclass[sigconf]{acmart}
\AtBeginDocument{%
  }

\setcopyright{acmlicensed}
\copyrightyear{2018}
\acmYear{2018}
\acmDOI{XXXXXXX.XXXXXXX}
\acmConference[Conference acronym 'XX]{Make sure to enter the correct
  conference title from your rights confirmation email}{June 03--05,
  2018}{Woodstock, NY}
\acmISBN{978-1-4503-XXXX-X/2018/06}

\usepackage{algorithm}
\usepackage{algorithmic}
\usepackage{booktabs}   
\usepackage{enumitem}
\usepackage{multirow}   
\usepackage{colortbl}   
\usepackage{xcolor}     
\usepackage{geometry}

\usepackage[most]{tcolorbox}
\definecolor{myblue}{RGB}{235, 245, 250}
\definecolor{mygreen}{RGB}{235, 250, 235}
\definecolor{myred}{RGB}{250, 235, 235}
\definecolor{mygray}{RGB}{245, 245, 245}

\definecolor{lightgray}{gray}{0.92}

\begin{document}

\title{Unlocking Biological Workflows for Robust Protein-Text Question Answering: A Dual-Dimensional RAG Framework}


\author{Li Ding}
\affiliation{
  \institution{Sichuan University}
  \country{China}}
  \email{dinglee@stu.scu.edu.cn}

\author{Duanyu Feng}
\affiliation{
  \institution{Sichuan University}
  \country{China}}
  \email{fengduanyuscu@stu.scu.edu.cn}

\author{Chen Huang}
\affiliation{
  \institution{National University of Singapore}
  \country{Singapore}}
  \email{huang\_chen@nus.edu.sg}

\author{Yangshuai Wang}
\affiliation{
  \institution{National University of Singapore}
  \country{Singapore}}
  \email{yswang@nus.edu.sg}

\author{Yang Li}
\affiliation{
  \institution{National University of Singapore}
  \country{Singapore}}
  \email{liyangum@nus.edu.sg}

\author{Wenqiang Lei}
\affiliation{
  \institution{Sichuan University}
  \country{China}}
  \email{wenqianglei@scu.edu.cn}

\author{See-Kiong Ng}
\affiliation{
  \institution{National University of Singapore}
  \country{Singapore}}
  \email{seekiong@nus.edu.sg}








\begin{abstract}
Protein-Text Question Answering (QA) is crucial for interpreting biological sequences through natural language. The integration of Large Language Models (LLMs) with Retrieval-Augmented Generation (RAG) that efficiently leverages biological databases and facilitates reasoning offers a potent approach for it.
However, constrained by the standard RAG pipeline, these models often rely on curated, static datasets instead of expert-proven biological workflows, lacking the fine-grained information processing and struggling to generalize to novel (OOD) proteins.
To bridge this gap, we propose 2D-ProteinRAG, a novel framework that empowers LLMs to operate within the gold-standard biological research workflow (BLAST). To further extract high-quality information from noisy retrieval contexts, we introduce a dual-dimensional (2D) filtering strategy following the expert analytical paradigms. Horizontal Fine-grained Attribute Alignment utilizes a lightweight, intent-aware discriminative filter to prune irrelevant metadata and align database entries with specific user queries. Vertical Homology-based Semantic Denoising resolves functional contradictions and redundancy across multiple homologs via hierarchical clustering.
Extensive evaluations on both In-Distribution and diverse biological OOD benchmarks demonstrate that 2D-ProteinRAG consistently achieves state-of-the-art performance,  outperforming fine-tuned baselines and other RAG methods. Our results validate the framework's robustness and scalability, providing a practical solution for interpreting protein functions in real-world scientific scenarios. Our code and demo will be available upon publication.

\end{abstract}



\received{20 February 2007}
\received[revised]{12 March 2009}
\received[accepted]{5 June 2009}

\maketitle

\section{Introduction}





Protein-Text Question Answering (QA) has emerged as a fundamental paradigm in AI-driven biological discovery, providing a natural language interface to decode the functional logic of amino acid sequences~\cite{liu-etal-2024-prott3, wu-etal-2025-rethinking-text, jararweh-etal-2025-protein2text}. By translating complex protein with specific user instruction into interpretable insights, such as molecular functions, catalytic activities, and subcellular localizations~\cite{fang2023mol, 10.1093/bib/bbl004}, Protein-Text QA bridges the semantic gap between biological "code" and human scientific knowledge. In practical workflows like drug discovery, this capability serves as a critical biological prior, enabling researchers to efficiently screen protein candidates through intuitive natural language queries~\cite{xiao2025proteingptmultimodalllmprotein}.



Recently, given their reasoning capabilities, Large Language Models (LLMs) have been established as the predominant framework for addressing this task~\cite{fan2025computationalproteinscienceera, xiao-etal-2025-protein, 10.1145/3726302.3730064}. 
However, since generic LLMs lack an inherent understanding of biological sequences, initial research focused on domain-specific fine-tuning or continued pre-training to bridge the semantic gap between amino acid sequences and natural language~\cite{10.5555/3618408.3620023, liu-etal-2024-prott3, jararweh-etal-2025-protein2text, 10979347, 10.1093/bioinformatics/btaf396}. 
To reduce these high computational costs and leverage the broad reasoning power of frozen LLMs, recent efforts have shifted toward Retrieval-Augmented Generation (RAG)~\cite{shaw2024protex, ma-etal-2024-retrieved, wu-etal-2025-rethinking-text}. 
These methods typically search for similar sequences within curated subsets of existing databases to provide the model with functional priors~\cite{wu-etal-2025-rethinking-text}. They construct the retrieval index using sequence-annotation pairs derived from fine-tuning datasets, which typically contain annotations for only a single or a limited set of attributes.
However, most existing protein-oriented RAG frameworks still remain shallow migrations of generic NLP architectures, lacking a deep integration of the reasoning used by biologists to solve similar problems~\cite{Whisstock2003-ch, Lee2007-fk, 10.1093/bib/bbl004}.  
This conceptual gap creates two primary bottlenecks. 
First, due to the inherent complexity of biological data, current models struggle to distill fine-grained, task-relevant evidence as an expert would, leading to the inclusion of significant retrieval noise within the context. 
Second, because the performance of these models is heavily constrained by the distributions of their training datasets, they remain vulnerable to performance collapse in Out-of-Distribution (OOD) scenarios.



To bridge this gap, we propose 2D-ProteinRAG, a novel framework that empowers LLMs to operate within the gold-standard biological research workflow. Unlike previous methods that rely on specialized, pre-curated datasets for RAG, our approach integrates homology search tools (BLAST) on UniProt database to query comprehensive databases, directly mimicking the practical workflows of biologists. This integration inherently resolves the OOD challenge by accessing the entire known biological universe. More importantly, to distill the fine-grained information and mitigate subsequent noise inherent in raw retrieval results, we introduce a dual-dimensional (2D) filtering strategy inspired by expert analytical paradigms:
(1) \textbf{Horizontally (Fine-grained Attribute Alignment)}: We introduce a lightweight, intent-aware discriminative filter to address the granularity mismatch between exhaustive database entries and specific user queries. Rather than relying on the generative model to sift through irrelevant text, we train a content-agnostic classifier via teacher-student distillation. This model learns to map instructions to relevant biological attribute tags (e.g., "Molecular Function" vs. "Subcellular Location"), effectively mimicking an expert’s ability to selectively identify task-specific evidence while pruning noisy, irrelevant metadata.
(2) \textbf{Vertically (Homology-based Semantic Denoising)}: We address the conflicting information inherent in raw homology retrieval. Grounded in the biological principle that "structure determines function,"  we employ hierarchical semantic clustering to group information into distinct functional clusters, and then anchor our selection to the cluster associated with the most structurally similar sequences. 



We evaluate 2D-ProteinRAG across extensive benchmarks, including both In-Distribution and Out-of-Distribution protein-text datasets. Experimental results demonstrate that our framework consistently achieves state-of-the-art performance, notably outperforming existing protein-oriented RAG models by 36 points in key metrics such as E-BLEU. To further validate this robustness, we conducted more tests under diverse biological OOD scenarios, e.g. Prot-Inst-OOD\cite{wu-etal-2025-rethinking-text}. Even in these challenging settings, our model demonstrates superior capability in capturing functional signals, yielding a 40 points improvement over the fine-tuned baselines. Our ablation studies also confirm the effectiveness of our dual-dimensional filtering strategy, showing that both the Horizontal Fine-grained Attribute Alignment and the Vertical Homology-based Semantic Denoising are essential for distilling high-fidelity knowledge from noisy retrieval contexts.

Our contributions are summarized as follows:
(1) We propose a novel paradigm that integrates gold-standard biological tools (BLAST) into the protein-text QA pipeline. By directly accessing comprehensive, raw biological databases rather than static curated subsets, our framework inherently overcomes the Out-of-Distribution (OOD) generalization limitations prevalent in existing models.
(2) To distill the fine-grained information in noisy retrieval, we design a 2D filtering strategy that simulates expert reasoning. Horizontally, an intent-aware, content-agnostic filter selectively prunes irrelevant attribute metadata to align with specific instructions. Vertically, a structure-anchored semantic clustering module resolves functional contradictions and eliminates redundancy across distant homologs.
(3) Extensive experiments on both In-Distribution and complex OOD benchmarks demonstrate that 2D-ProteinRAG consistently achieves state-of-the-art performance. Our results validate the framework's robustness in extracting high quality biological information, offering a practical and scalable solution for real-world protein analysis.

\section{Related work}

\textbf{LLM-based Approaches for Protein-Text QA.}
Recent research has extensively explored Large Language Models (LLMs) to bridge the semantic gap between protein sequences and natural language ~\cite{xiao-etal-2025-protein}. These approaches can be broadly categorized into parametric adaptation and retrieval-augmented generation.
For the parametric adaptation, early efforts focused on direct fine-tuning or multi-modal alignment. The first category treats protein sequences as a specialized language, fine-tuning base LLMs on protein-text pairs to learn biological semantics ~\cite{10979347, taylor2022galacticalargelanguagemodel}. The second category views proteins as a distinct modality, employing separate encoders (e.g., ESM~\cite{doi:10.1126/science.ade2574, doi:10.1126/science.ads0018}) aligned with LLMs via adapters to facilitate cross-modal understanding ~\cite{10.5555/3618408.3620023, liu-etal-2024-prott3, jararweh-etal-2025-protein2text}. However, these "model-centric" approaches are not only computationally expensive but also suffer from catastrophic forgetting, where the adaptation to protein syntax degrades the model's general reasoning capabilities~\cite{doi:10.1073/pnas.1611835114}.
Therefore, to mitigate the knowledge obsolescence of parametric models, retrieval-based methods have gained traction by fetching relevant context from external knowledge bases~\cite{shaw2024protex, wu-etal-2025-rethinking-text}. For instance, RAPM~\cite{wu-etal-2025-rethinking-text} utilizes a dense retriever to fetch property-oriented descriptions from a curated database. ProtEx\cite{shaw2024protex} trains a Transformer model on curated exemplars retrieved via homology-based sequence alignment within on label classification tasks.
Despite their promising results, the current protein-oriented RAG frameworks primarily represent shallow migrations of generic NLP architectures. These methods typically rely on simplistic similarity-based retrieval from static, curated datasets, which inherently restricts their coverage and renders them vulnerable to Out-of-Distribution (OOD) challenges. More critically, they lack fine-grained consideration of biological attributes and the ability to denoise complex biological information, leading to suboptimal performance in real-world scenarios.

\textbf{Bioinformatics-Native Workflows for Protein Analysis.}
In contrast to AI-centric paradigms, traditional bioinformatics relies on canonical workflows rooted in evolutionary biology. The fundamental principle is that evolutionary conservation implies functional similarity ~\cite{Whisstock2003-ch}.
The de facto standard for function prediction involves querying comprehensive, raw databases (e.g., UniProtKB ~\cite{10.1093/nar/gkae1010}) via homology search tools like BLAST~\cite{Lee2007-fk}. Unlike the static, vector-based retrieval used in standard RAG, BLAST identifies evolutionary relatives by calculating sequence alignments and E-values, providing a highly interpretable and comprehensive set of biological evidence. This approach allows researchers to access the most up-to-date biological knowledge without being restricted by the distribution of a specific training set.
Critically, the raw output from a homology search is not a direct answer but a collection of exhaustive metadata from multiple homologs \cite{Szafron2004-ra}. Biologists perform a sophisticated information processing task to distill this data. They must selectively focus on task-relevant attributes (e.g., specific catalytic sites) while filtering out irrelevant annotations, and resolve functional contradictions that often arise from distant homologs ~\cite{Rost2003-jm, Devos2000-ka}. This expert reasoning, moving from a "noisy set of relatives" to a "precise functional consensus", is the key to accurate biological discovery. Motivated by this interpretable paradigm, our work aims to integrate the BLAST-based workflow directly into the LLM reasoning loop, specifically mimicking the expert logic of fine-grained attribute filtering and homology-based denoising.



\begin{figure*}[h]
  \centering
  \includegraphics[width=0.9\linewidth]{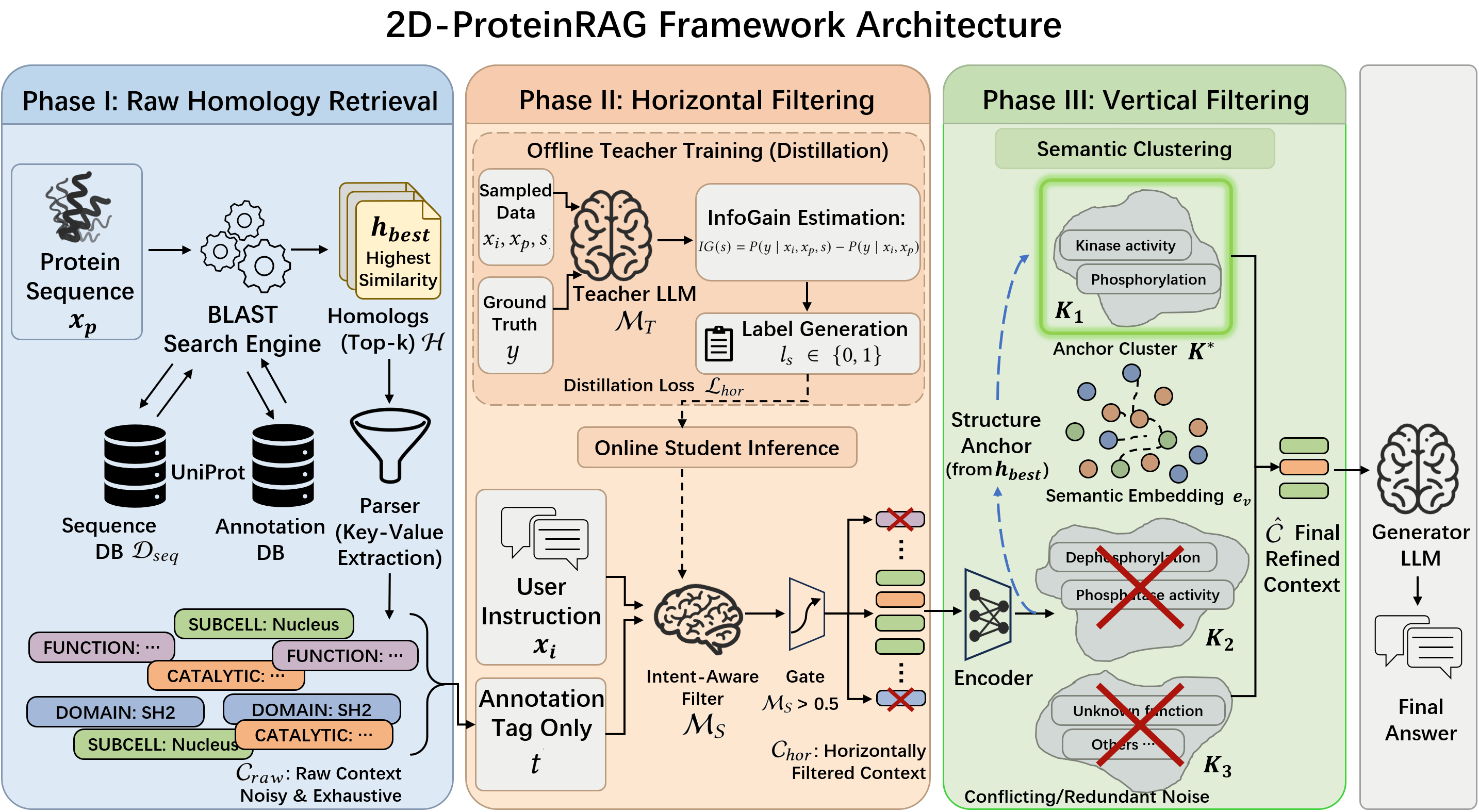}
  \setlength{\abovecaptionskip}{0pt}   
\setlength{\belowcaptionskip}{0pt}
  \caption{Overview of the 2D-ProteinRAG Framework. The workflow consists of three phases (1) Raw Homology Retrieval: The system employs BLAST to retrieve homologous sequences from UniProt. (2) Horizontal Filtering: An intent-aware filter selects relevant attribute tags based on the user's specific instruction. (3) Vertical Filtering: The remaining descriptions are grouped via DBSCAN. The cluster consistent with the highest-similarity homolog (Structure Anchor) is retained as the valid context $\hat{C}$.}
  \Description[Overview of the 2D-ProteinRAG Framework.]{The workflow consists of three phases (1) Raw Homology Retrieval: The system employs BLAST to retrieve homologous sequences from UniProt. (2) Horizontal Filtering: An intent-aware filter selects relevant attribute tags based on the user's specific instruction. (3) Vertical Filtering: The remaining descriptions are grouped via DBSCAN. The cluster consistent with the highest-similarity homolog (Structure Anchor) is retained as the valid context $\hat{C}$.}
  \label{sec3}
\end{figure*}

\section{Problem Formulation}

\textbf{Task Definition.} Let $x_p$ denote a protein sequence represented as an amino acid string, and $x_i$ be a natural language instruction specifying a biological query. The goal of \textbf{Protein-Text QA} is to generate a textual response $y$ that accurately addresses $x_i$ based on the biological properties of $x_p$. In our retrieval-augmented setting, the model accesses a comprehensive biological database $\mathcal{D} = \{ (h_j, a_j) \}$, where $h_j$ and $a_j$ represent homologous sequences and their exhaustive functional information/annotations, respectively. The objective is to generate the response $y$ from a generative model $\mathcal{M}$ conditioned on the input pair and the retrieved evidence: $y = \mathcal{M}(x_p, x_i, \mathcal{D})$. The core challenge is to distill task-specific, high-fidelity signals from the noisy and multi-faceted metadata $a_j$ in $\mathcal{D}$ to align with the specific intent of $x_i$.

\section{Method}


The 2D-ProteinRAG framework is designed to generate biologically accurate responses by distilling task-relevant knowledge from the noisy outputs of standard homology search pipelines. As illustrated in Figure \ref{sec3}, The pipeline is structured into two major components: standard Raw Homology Retrieval (Phase I) and our proposed Dual-Dimensional (2D) Filtering Strategy (Phases II \& III).
\begin{itemize}[leftmargin=*]

\item \textbf{Phase I: Raw Homology Retrieval (The Standard Workflow).} To ensure biological reliability and overcome OOD limitations, we first integrate the gold-standard \textbf{BLAST} \cite{Altschul1990-fr} tool to identify evolutionary relatives of the query protein $x_p$. Exhaustive annotations/information $\{a_j\}$ are fetched from UniProt \cite{10.1093/nar/gkae1010} and parsed into attribute-value snippets, forming a comprehensive but noisy evidence pool 
$\mathcal{C}_{raw}$.

\item \textbf{Phase II: Horizontal Filtering (Fine-grained Attribute Alignment).} As the first dimension of our 2D strategy, we address the granularity mismatch by training an intent-aware filter. This module selectively retains only the attribute tags (e.g., \texttt{FUNCTION}) that align with the specific user instruction $x_i$, distilling the encyclopedic metadata into an attribute-aligned context $\mathcal{C}_{hor}$ in the information $\{a_j\}$.

\item \textbf{Phase III: Vertical Filtering (Homology-based Semantic Denoising).} As the second dimension, we address semantic conflicts across different homologs. By performing hierarchical clustering on the descriptions and anchoring the selection to the cluster most structurally consistent with $x_p$, we eliminate redundancy and noise to produce the final high-fidelity context $\hat{\mathcal{C}}$ from the information $\{a_j\}$.

\end{itemize}

\subsection{Phase I: Raw Homology Retrieval}

Given an input protein sequence $x_p$ and a user instruction $x_i$, our objective is to ground the reasoning of the model $\mathcal{M}$ in verified biological evidence. Following the standard bioinformatics protocol, we first utilize the \textbf{BLAST} algorithm to query the comprehensive biological database $\mathcal{D}$. This step identifies a set of $k$ homologous sequences sorted by their alignment significance (e.g., E-values):
\begin{equation}
    \mathcal{H} = \{h_1, h_2, \dots, h_k\} = \text{BLAST}(x_p, \mathcal{D}).
\end{equation}

For each homolog $h_j \in \mathcal{H}$, we retrieve its corresponding expert-curated functional annotations. Since biological metadata in databases like UniProt \cite{10.1093/nar/gkae1010} is inherently structured, we parse the raw entry of each homolog into a set of fine-grained snippets represented as attribute-value pairs:
\begin{equation}
    C_j = \{ (t_{j,m}, v_{j,m}) \}_{m=1}^{M_j},
\end{equation}
where $t_{j,m}$ denotes a specific attribute tag (e.g., \texttt{FUNCTION}, \texttt{SUBCELLULAR LOCATION}) and $v_{j,m}$ represents the associated textual description. Finally, we define the initial raw context $\mathcal{C}_{raw}$ as the union of all snippets retrieved from the top-$k$ homologs:
\begin{equation}
    \mathcal{C}_{raw} = \bigcup_{j=1}^k C_j.
\end{equation}
This exhaustive knowledge pool $\mathcal{C}_{raw}$ serves as the primary evidence source for our subsequent dual-dimensional filtering strategy, capturing the broad biological context of $x_p$ while inheriting the noise and redundancy of raw retrieval results.



\subsection{Phase II: Horizontal Filtering (Fine-grained Attribute Alignment)}

Standard biological databases contain exhaustive attributes, most of which are irrelevant to specific user instructions. To address this, we employ a Horizontal Filtering strategy driven by a teacher-student distillation framework. This module trains a lightweight, intent-aware filter to prune irrelevant metadata, ensuring that only instruction-aligned attributes are retained.

\textbf{Information Gain Estimation (Teacher).} 
We utilize a capable LLM as the teacher model $\mathcal{M}_T$ to quantify the utility of each attribute via Information Gain (IG). Given an instruction-sequence pair $(x_i, x_p)$ and its ground truth answer $y$, we consider each snippet $s = (t, v) \in \mathcal{C}_{raw}$ as a candidate evidence unit. 
To address the partial relevance issue in multi-attribute tasks, we propose a Segment-wise Information Gain metric. Let the ground truth answer $y$ be decomposed into a sequence of $M$ semantic fragments $S_y = \{s_1, s_2, \dots, s_M\}$ (e.g., split by sentences). We define the effective information gain of a fine-grained annotation $d$ as the maximum gain it provides to any individual fragment $s_k$:$$\text{IG}_{\text{seg}}(d|x) = \max_{k \in \{1, \dots, M\}} \left( p_{\phi}(s_k|x, d) - p_{\phi}(s_k|x) \right),$$where $s_k$ denotes the $k$-th fragment of the ground truth $y$. $p_{\phi}(s_k|x, d)$ is the generation probability of fragment $s_k$ given the query $x$ and document $d$. $p_{\phi}(s_k|x)$ is the base generation probability of fragment $s_k$ given only the query. By maximizing over fragments, we ensure that an annotation $d$ is rewarded if it accurately aligns with any specific attribute described in $y$, regardless of its relevance to the remaining parts of the answer. Based on this metric, we assign a binary relevance label $l_s \in \{0, 1\}$ to each snippet: $l_s = 1$ if $IG(s) > \tau$ (where $\tau$ is a significance threshold), and $0$ otherwise. 

\textbf{Intent-Aware Filter Training (Student).} 
To ensure inference efficiency and zero-shot generalization to unseen protein content, we train a lightweight student model $\mathcal{M}_S$ (e.g., a RoBERTa-based classifier) to predict attribute relevance. Critically, to force the model to learn the semantic alignment between the \textit{User Intent} and the \textit{Database Schema} rather than overfitting to specific textual values, the input to $\mathcal{M}_S$ is restricted to the instruction $x_i$ and the attribute tag $t$. The model is optimized by minimizing the binary cross-entropy (BCE) loss:
\begin{equation}
    \mathcal{L}_{hor} = - \sum_{(x_i, t) \in \mathcal{D}_{train}} [l_s \log \mathcal{M}_S(x_i, t) + (1-l_s) \log (1 - \mathcal{M}_S(x_i, t))],
\end{equation}
where $\mathcal{M}_S(x_i, t)$ represents the predicted probability that tag $t$ is relevant to instruction $x_i$.

\textbf{Inference.} 
During the inference phase, the trained filter acts as a content-agnostic gatekeeper. For a retrieved set $\mathcal{C}_{raw}$, we evaluate each attribute tag $t$ and retain only those snippets with high relevance scores. The resulting horizontally aligned context is defined as follows:
\begin{equation}
    \mathcal{C}_{hor} = \{ (t, v) \in \mathcal{C}_{raw} \mid \mathcal{M}_S(x_i, t) > 0.5 \}.
\end{equation}
This process effectively distills the encyclopedic metadata into a compact, intent-aligned subset, significantly reducing noise before the semantic analysis in Phase III.

\subsection{Phase III: Vertical Filtering (Homology-based Semantic Denoising)}

While Horizontal Filtering aligns the context with user intent, the resulting set $\mathcal{C}_{hor}$ may still contain conflicting evidence derived from evolutionarily distant homologs (e.g., distinct functional divergences). To resolve these contradictions, we propose a Vertical Filtering strategy grounded in the biological axiom that \textit{"structure determines function."} This phase employs a cluster-and-anchor mechanism to synthesize a consistent biological consensus.

\textbf{Semantic Clustering.} 
First, to handle redundancy and group semantically similar evidence, we extract all textual values $\mathcal{V} = \{ v \mid (t, v) \in \mathcal{C}_{hor} \}$ from the aligned snippets. We utilize a pre-trained encoder (e.g., RoBERTa) to map each description $v$ into a dense semantic vector $\mathbf{e}_v$. 
Subsequently, we apply Density-Based Spatial Clustering of Applications with Noise (DBSCAN) on these embeddings to partition the evidence into $N$ distinct semantic clusters \footnote{According to the DBSCAN, the $N$ can be automatically chosen.}:
\begin{equation}
    \mathcal{K} = \{K_1, K_2, \dots, K_N\} = \text{DBSCAN}(\{ \mathbf{e}_v \}_{v \in \mathcal{V}}).
\end{equation}
Descriptions falling within the same cluster $K_n$ are treated as representing the same underlying biological fact, effectively consolidating redundant annotations.

\textbf{Structure-Driven Anchoring.} 
Clusters may still represent conflicting functions (e.g., one cluster suggests "Kinase activity" while another suggests "Phosphatase activity"). To resolve this, we prioritize evidence from the homolog most structurally similar to the query protein $x_p$.
Let $h_{best} = h_1$ be the top-ranked homolog from the BLAST retrieval in Phase I, which serves as our \textbf{Structural Anchor}. We identify the optimal cluster $K^*$ that contains the evidence derived from this anchor:
\begin{equation}
    K^* = \{ K_n \in \mathcal{K} \mid \exists v \in K_n \text{ derived from } h_{best} \}.
    \label{eq:h}
\end{equation}
By selecting the cluster $K^*$, we effectively perform a soft vote weighted by structural similarity, discarding outliers and noise introduced by distant homologs.

\textbf{Final Context Generation.} 
The final high-fidelity context $\hat{\mathcal{C}}$ from $K^*$ is constructed by aggregating the descriptions from the selected anchor cluster. This context is then concatenated with the user instruction $x_i$ and the protein sequence $x_p$ to prompt the LLM for the final generation:
\begin{equation}
    y = \mathcal{M}(x_p, x_i, \hat{\mathcal{C}}).
\end{equation}
This dual-filtered context ensures that the generative model is grounded in evidence that is both intent-relevant (Horizontal) and biologically consistent (Vertical). \footnote{A pseudocode of the 2D-ProteinRAG can be found in Appendix \ref{app:Pseudocode}.}

\section{Experiments}

In this section, we present a comprehensive evaluation of the 2D-ProteinRAG framework. To validate its effectiveness in mitigating the noise inherent in raw biological retrieval and overcoming OOD generalization bottlenecks, we benchmark our approach against a diverse set of baselines across two datasets and four distinct tasks. Our experiments are designed to answer three core research questions:
(1)
\textbf{RQ1 (Generalization):} Can 2D-ProteinRAG mitigate the performance collapse often observed in traditional methods on OOD data?
(2) \textbf{RQ2 (Robustness):} How does the model perform under varying degrees of sequence homology constraints, representing broader and more challenging generalization scenarios?
(3) \textbf{RQ3 (Denoising):} Does the dual-dimensional filtering strategy effectively distill high-fidelity evidence from noisy raw retrieval contexts?


\begin{table*}[htbp]
\centering
\small 
\setlength{\abovecaptionskip}{0pt}   
\setlength{\belowcaptionskip}{0pt}
\caption{Main results on Mol-Instructions dataset. We compare our \textbf{2D-ProteinRAG} with various baselines. The best results are highlighted in \textbf{bold}. The best results are highlighted in \textbf{bold}, and the second best results are \underline{underlined}.}
\vspace{2mm}
\label{tab:main_results1}
\normalsize
    \resizebox{\textwidth}{!}{
    \setlength{\tabcolsep}{1pt}
    \footnotesize
\begin{tabular}{l c cccc cccc cccc cccc}
\toprule
\multirow{2}{*}{\textbf{Model}} & \multirow{2}{*}{\textbf{Param}} & \multicolumn{4}{c}{\textbf{General Description}} & \multicolumn{4}{c}{\textbf{Protein Function}} & \multicolumn{4}{c}{\textbf{Catalytic Activity}} & \multicolumn{4}{c}{\textbf{Domain/Motif}} \\
\cmidrule(lr){3-6} \cmidrule(lr){7-10} \cmidrule(lr){11-14} \cmidrule(lr){15-18}
 & & \textbf{E-BL2} & \textbf{E-BL4} & \textbf{BL4} & \textbf{RG-L} & \textbf{E-BL2} & \textbf{E-BL4} & \textbf{BL4} & \textbf{RG-L} & \textbf{E-BL2} & \textbf{E-BL4} & \textbf{BL4} & \textbf{RG-L} & \textbf{E-BL2} & \textbf{E-BL4} & \textbf{BL4} & \textbf{RG-L} \\
\midrule

\rowcolor{lightgray} \multicolumn{18}{l}{\textit{\textbf{Fine-tuned LLMs}}} \\
ProLLama     & 8B   & 20.2  & 18.9  & 32.6  & 44.5  & 22.0  & 20.4  & 34.0  & 46.4  & 20.2  & 19.5  & 38.4  & 52.5  & 37.7  & 36.7  & 39.1  & 49.3 \\
Galactica    & 1.3B  & 55.8  & 54.5  & \underline{57.4}  & \textbf{63.0}  & 49.3  & 47.5  & \textbf{47.3}  & \textbf{55.2}  & 67.7  & 67.1  & \textbf{62.3}  & \textbf{66.6}  & \textbf{69.6}  & \textbf{68.4}  & \textbf{45.7}  & \textbf{52.0} \\
BioGPT       & 347M  & 46.4  & 44.6  & 45.5  & 54.5  & 40.3  & 38.8  & 42.3  & 51.8  & 54.9  & 54.0  & 46.0  & 62.4  & 60.6  & 59.4  & 42.0  & 50.9 \\
Llama-3.1     & 8B  & 49.2  & 47.9  & 52.9  & 59.5  & 43.1  & 41.4  & 44.0  & \underline{52.8}  & 60.8  & 60.2  & \underline{59.1}  & \underline{64.8}  & \underline{64.9}  & \underline{63.7}  & \underline{45.2}  & \underline{52.0} \\
Protein2Text & 8.65B & 27.6  & 13.0  & 2.8  & 10.7  & 45.1  & 33.4  & 6.4  & 15.1  & 32.2  & 22.9  & 4.7  & 11.4  & 40.7  & 32.6  & 2.8  & 10.5 \\

\rowcolor{lightgray} \multicolumn{18}{l}{\textit{\textbf{Few-shot prompted Models}}} \\
Llama-3.1$_{w/ Fewshot}$       & 8B & 0.1  & 0.0  & 2.2  & 16.2  & 0.0  & 0.0  & 9.9  & 24.6  & 0.0  & 0.0  & 4.3  & 19.3  & 0.0  & 0.0  & 2.1  & 15.1 \\
Llama-3.3$_{w/ Fewshot}$       & 70B & 0.3  & 0.1  & 1.9  & 16.3  & 0.1  & 0.0  & 8.2  & 23.3  & 0.0  & 0.0  & 6.2  & 21.7  & 0.0  & 0.0  & 0.7  & 11.5 \\
Gemini-3-Flash$_{w/ Fewshot}$  & N/A  & 1.6  & 0.9  & 5.4  & 17.8  & 0.5  & 0.3  & 4.9  & 18.7  & 0.8  & 0.6  & 11.0  & 29.7  & 0.5  & 0.4  & 6.2  & 25.2 \\

\rowcolor{lightgray} \multicolumn{18}{l}{\textit{\textbf{RAG prompted Models}}} \\
Llama-3.1$_{w/ BLAST}$       & 8B & 53.6  & 51.9  & 46.9  & 56.1  & 12.0  & 8.7  & 12.0  & 20.9  & 17.6  & 14.2  & 15.1  & 21.8  & 2.7  & 1.8  & 1.6  & 8.6 \\
Llama-3.3$_{w/ BLAST}$       & 70B & 44.2  & 39.7  & 34.9  & 40.4  & 50.9  & 43.4  & 21.8  & 32.5  & 60.0  & 54.5  & 25.3  & 36.0  & 29.2  & 22.0  & 3.3  & 11.7 \\
Gemini-3-Flash$_{w/ BLAST}$  & N/A  & 53.7  & 45.3  & 25.8  & 35.6  & 46.9  & 35.4  & 12.9  & 24.2  & 64.0  & 57.7  & 27.9  & 44.4  & 46.3  & 39.4  & 14.5  & 36.1 \\
Llama-3.1$_{w/ RAPM}$       & 8B & 52.4  & 50.8  & 49.7  & 54.4  & 47.3  & 43.7  & 31.6  & 39.1  & 61.3  & 59.4  & 48.8  & 53.9  & 17.4  & 15.4  & 7.2  & 20.0  \\
Llama-3.3$_{w/ RAPM}$       & 70B & 32.7  & 29.3  & 32.2  & 41.1  & \underline{56.8}  & 51.8  & \underline{35.0}  & 41.8  & \textbf{83.8}  & \textbf{82.3}  & 44.8  & 54.1  & 40.9  & 36.5  & 10.7  & 25.5 \\
Gemini-3-Flash$_{w/ RAPM}$  & N/A  & 51.0  & 49.1  & 47.8  & 56.3  & 56.8  & \underline{51.9}  & 32.1  & 40.8  & 62.8  & 60.3  & 47.1  & 56.5  & 50.6  & 46.8  & 21.8  & 40.2  \\

\rowcolor{lightgray} \multicolumn{18}{l}{\textit{\textbf{Our method}}} \\
Llama-3.1$_{w/ 2D-ProteinRAG}$       & 8B & \textbf{68.7}  & \textbf{67.3}  & \textbf{58.8}  & \underline{62.9}  & 40.0  & 35.4  & 29.3  & 36.3  & 72.3  & 66.3  & 45.2  & 60.5  & 27.4  & 24.1  & 11.7  & 26.1 \\
Llama-3.3$_{w/ 2D-ProteinRAG}$       & 70B & 45.9  & 42.7  & 37.6  & 53.0  & \textbf{60.8}  & \textbf{54.6}  & 33.8  & 42.9  & 68.6  & 63.0  & 40.2  & 53.7  & 37.4  & 31.2  & 7.6  & 18.2 \\
Gemini-3-Flash$_{w/ 2D-ProteinRAG}$  & N/A  & \underline{59.8}  & \underline{54.7}  & 47.0  & 56.1  & 54.8  & 44.0  & 19.6  & 30.8  & \underline{73.8}  & \underline{67.5}  & 44.7  & 61.4  & 45.4  & 40.1  & 17.6  & 39.0 \\

\bottomrule
\end{tabular}}
\end{table*}

\subsection{Experimental Setup}

\textbf{Datasets.} To provide a rigorous evaluation across both In-Distributi-on (ID) and Out-of-Distribution (OOD) scenarios, we conduct experiments on two datasets covering four core tasks: General Description Generation, Protein Function Prediction, Catalytic Activity Prediction, and Domain/Motif Prediction. For the ID benchmark, we utilize the protein-oriented subset of Mol-Instructions \cite{fang2023mol}, a standard collection of expert-curated sequence-instruction pairs used to assess fundamental model capabilities on well-aligned data. To further stress-test generalization, we employ Prot-Inst-OOD \cite{wu-etal-2025-rethinking-text}, a benchmark specifically constructed to eliminate the data leakage prevalent in existing datasets. By enforcing strict constraints where test proteins share minimal sequence homology with the training corpus, Prot-Inst-OOD serves as the primary testbed for verifying whether the training module of our framework can leverage biological workflows to generalize to novel, unseen proteins. \footnote{More details of using these models can be found in Appendix \ref{app:data}.} For the retrieval component, we employ Swiss-Prot from UniProtKB~\cite{10.1093/nar/gkae1010} as our target knowledge base, ensuring access to gold-standard functional annotations.

\textbf{Baselines.} To rigorously benchmark the effectiveness of 2D-ProteinRAG, we compare it against three distinct categories of methods, ranging from domain-specific fine-tuning to existing retrieval-augmented paradigms:
(1) \textbf{Fine-tuned LLMs (Parametric Knowledge):} We evaluate models that rely solely on internalized knowledge acquired through weight updates. This category includes domain-specialized experts, BioGPT (347M)~\cite{10.1093/bib/bbac409}, Galactica (1.3B)~\cite{taylor2022galacticalargelanguagemodel}, Protein2Text (8.65B)~\cite{jararweh-etal-2025-protein2text}, and ProLLama (8B)~\cite{10979347}, as well as the general-purpose Llama-3.1 (8B)~\cite{grattafiori2024llama3herdmodels} fine-tuned on our training data to represent standard domain adaptation.
(2) \textbf{Few-shot Prompted Models (In-Context Reasoning):} To assess the inherent reasoning capabilities of frozen LLMs without external retrieval, we benchmark Llama-3.1 (8B)~\cite{grattafiori2024llama3herdmodels}, the large-scale Llama-3.3 (70B)~\cite{grattafiori2024llama3herdmodels}, and Gemini-Flash~\cite{comanici2025gemini25pushingfrontier} using few-shot prompting. This sets the performance floor for general-purpose reasoners.
(3) \textbf{RAG Baselines (Retrieval-Augmented):} We equip the base LLMs (Llama-3.1 (8B), Llama-3.3 (70B), and Gemini-Flash) with two representative retrieval strategies: BLAST (Naive Retrieval) \cite{Altschul1990-fr}, which provides the model with raw, unfiltered homology search results. This baseline is critical for demonstrating the challenge of retrieval noise and the necessity of our denoising strategy; RAPM \cite{wu-etal-2025-rethinking-text}, the current state-of-the-art protein RAG framework that utilizes dense retrieval on curated knowledge bases. \footnote{More details of the construction of these models can be found in Appendix \ref{app:hyp}.}

\textbf{Evaluation Metrics.} Following the standard benchmark~\cite{fang2023mol, liu-etal-2024-prott3, 10.1093/bioinformatics/btaf396, fei2025prottextv, wu-etal-2025-rethinking-text}, we employ a multi-dimensional evaluation protocol. First, we primarily utilize E-BLEU (including E-BL2, E-BL4)as our main metric. Unlike standard overlap metrics, E-BLEU specifically evaluates the accuracy of key biological entities (e.g., functional terms, molecule names) within the generated text~\cite{wu-etal-2025-rethinking-text}. This ensures that our evaluation captures scientific factual precision rather than just linguistic surface resemblance. Complementing this, we report standard NLP metrics including BLEU-4 and ROUGE-L to measure n-gram precision and recall based on the longest common subsequence, respectively, assessing the general fluency and structural coherence of the responses.

\textbf{Implementation details.}
For the retrieval phase of 2D-Protein-RAG framework, we employed the standard NCBI BLAST to query the database, retrieving the top-3. \textbf{To prevent data leakage, we excluded retrieved sequences that are identical to the query sequence, along with their associated annotations.} In the Horizontal Filtering module, the student model was initialized with Roberta \cite{liu2019robertarobustlyoptimizedbert} and fine-tuned for 4 epochs using the AdamW optimizer with a learning rate of 1e-5 and a batch size of 64. For Vertical Filtering, we utilized Roberta to generate semantic embeddings. \footnote{More details can be found in Appenidx \ref{app:imp}.}


\subsection{Main Results}

The comparative results on the Mol-Instructions (ID) and Prot-Inst-OOD (OOD) benchmarks are presented in Table \ref{tab:main_results1} and Table \ref{tab:main_results2}. Our analysis focuses on three key observations:


\textbf{2D-ProteinRAG achieves State-of-the-Art performance across the majority of tasks, whereas competing baselines may suffer from Data Leakage or OOD risks.}
First and foremost, 2D-ProteinRAG achieves dominant performance across the majority of tasks, establishing a new state-of-the-art standard. On the Mol-Instructions dataset, our framework utilizing Llama-3.1-8B yields the highest E-BLEU4 scores in General Description (67.3) and Protein Function (35.4), significantly outperforming both fine-tuned experts and RAPM. This dominance is further amplified on the challenging Prot-Inst-OOD benchmark, where 2D-ProteinRAG gets highest E-BLEU scores on all four tasks—achieving, for instance, a score of 72.7 in Catalytic Activity.
We also observe that in specific tasks like Catalytic Activity, the RAPM baseline (Llama-3.3) achieves exceptionally high E-BLEU4 scores (82.3), surpassing our method. However, this superiority is not stable, as its performance precipitously drops to 18.5 on the corresponding OOD task. We attribute this to the severe data leakage inherent in Mol-Instructions, where over 90\% of test samples are retrievable from the training set. This allows retrieval-based baselines trained on specific curated sets (RAPM) or fine-tuned models to "memorize" answers via similarity search or weight updates.


\textbf{2D-ProteinRAG resolves the generalization bottleneck, preventing the collapse observed in traditional methods on OOD data.}
As evidenced in Table \ref{tab:main_results2}, the Prot-Inst-OOD benchmark poses a severe challenge by enforcing strict homology constraints. Under these conditions, Fine-tuned LLMs experience a total failure; for instance, the Llama-3.1 model's score on General Description plummets to a negligible 0.4 of E-BLEU2, indicating that parametric knowledge fails to generalize to novel proteins. Similarly, standard RAG methods like RAPM struggle due to the limited coverage of their curated indices, scoring only 6.7 of E-BLEU2. 
In stark contrast, 2D-ProteinRAG maintains robust performance, achieving that all four tasks with scores higher than the baselines (e.g., 44.3 of E-BLEU2 in General Description). This qualitative leap confirms that integrating standard biological workflows (BLAST) allows our model to access the "open-world" biological evidence required for handling unseen sequences, a capability that closed-set training methods fundamentally lack. Furthermore, although our framework involves a trained discriminative filter, the learning process is intentionally designed to be content-agnostic. By focusing on the semantic alignment between user intent and database tags rather than memorizing specific protein-function pairs, our training remains inherently decoupled from the biological distribution, ensuring robust reliability even in extreme OOD scenarios.

\begin{table*}[htbp]
\centering
\small 
\setlength{\abovecaptionskip}{0pt}   
\setlength{\belowcaptionskip}{0pt}
\caption{Main results on Prot-Inst-OOD. The best results are highlighted in \textbf{bold}, and the second best results are \underline{underlined}.}
\vspace{2mm}
\label{tab:main_results2}
\normalsize
    \resizebox{\textwidth}{!}{
    \setlength{\tabcolsep}{1pt}
    \footnotesize
\begin{tabular}{l c cccc cccc cccc cccc}
\toprule
\multirow{2}{*}{\textbf{Model}} & \multirow{2}{*}{\textbf{Param}} & \multicolumn{4}{c}{\textbf{General Description}} & \multicolumn{4}{c}{\textbf{Protein Function}} & \multicolumn{4}{c}{\textbf{Catalytic Activity}} & \multicolumn{4}{c}{\textbf{Domain/Motif}} \\
\cmidrule(lr){3-6} \cmidrule(lr){7-10} \cmidrule(lr){11-14} \cmidrule(lr){15-18}
 & & \textbf{E-BL2} & \textbf{E-BL4} & \textbf{BL4} & \textbf{RG-L} & \textbf{E-BL2} & \textbf{E-BL4} & \textbf{BL4} & \textbf{RG-L} & \textbf{E-BL2} & \textbf{E-BL4} & \textbf{BL4} & \textbf{RG-L} & \textbf{E-BL2} & \textbf{E-BL4} & \textbf{BL4} & \textbf{RG-L} \\
\midrule

\rowcolor{lightgray} \multicolumn{18}{l}{\textit{\textbf{Fine-tuned LLMs}}} \\
ProLLama     & 8B   & 0.1  & 0.1  & 7.8  & 29.6  & 2.0  & 1.5  & 19.3  & 38.7  & 0.0  & 0.0  & 17.8  & 41.9  & 0.2  & 0.2  & 22.8  & 41.1 \\
Galactica    & 1.3B & 0.8  & 0.6  & 11.2  & 31.5  & 9.3  & 7.9  & \underline{25.3}  & \textbf{42.6}  & 1.4  & 1.2  & 22.8  & 44.1  & 4.6  & 4.2  & \textbf{26.4}  & \underline{42.5} \\
BioGPT       & 347M & 1.3  & 0.8  & 10.6  & 28.3  & 5.9  & 4.9  & 23.1  & 41.3  & 0.9  & 0.7  & 17.9  & 42.2  & 1.5  & 1.3  & 24.0  & 41.6 \\
Llama-3.1     & 8B  & 0.4  & 0.2  & 9.7  & 31.1  & 6.3  & 5.2  & 23.4  & \underline{41.5}  & 0.5  & 0.4  & 20.7  & 43.0  & 2.6  & 2.4  & \underline{25.5}  & \textbf{42.8} \\
Protein2Text & 8.65B & 9.3  & 22.7  & 2.3  & 10.0  & 26.2  & \underline{38.7}  & 5.2  & 13.5  & 11.0  & 17.8  & 3.2  & 9.1  & 22.2  & 30.4  & 2.0  & 7.8 \\

\rowcolor{lightgray} \multicolumn{18}{l}{\textit{\textbf{Few-shot prompted Models}}} \\
Llama-3.1$_{w/ Fewshot}$       & 8B & 0.1  & 0.0  & 2.2  & 15.9  & 0.1  & 0.0  & 9.8  & 24.2  & 0.0  & 0.0  & 4.0  & 19.0  & 0.0  & 0.0  & 2.2  & 15.6 \\
Llama-3.3$_{w/ Fewshot}$       & 70B & 0.2  & 0.0  & 1.9  & 16.1  & 0.2  & 0.1  & 7.9  & 22.9  & 0.0  & 0.0  & 5.8  & 21.3  & 0.0  & 0.0  & 0.6  & 11.5 \\
Gemini-3-Flash$_{w/ Fewshot}$  & N/A  & 0.7  & 0.3  & 4.2  & 16.5  & 0.7  & 0.4  & 4.5  & 18.5  & 0.6  & 0.4  & 10.8  & 29.3  & 0.3  & 0.2  & 6.4  & 25.1 \\

\rowcolor{lightgray} \multicolumn{18}{l}{\textit{\textbf{RAG prompted Models}}} \\
Llama-3.1$_{w/ BLAST}$       & 8B & 29.8  & 28.2  & \underline{31.6}  & 41.7  & 8.4  & 5.8  & 9.2  & 19.3  & 14.6  & 12.1  & 13.6  & 21.0  & 0.5  & 0.3  & 0.9  & 8.1 \\
Llama-3.3$_{w/ BLAST}$       & 70B & 21.9  & 18.3  & 21.7  & 30.4  & 40.3  & 32.4  & 17.5  & 29.1  & 63.7  & 57.4  & 24.6  & 35.5  & 28.9  & 21.7  & 3.2  & 12.3 \\
Gemini-3-Flas$h_{w/ BLAST} $ & N/A  & 35.3  & 27.6  & 19.2  & 28.8  & 38.0  & 27.1  & 10.2  & 22.5  & 65.6  & 58.7  & 26.5  & 43.3  & \underline{47.1}  & \underline{37.5}  & 13.4  & 35.5 \\
Llama-3.1$_{w/ RAPM}$       & 8B & 6.7  & 5.6  & 16.1  & 25.9  & 23.0  & 19.6  & 18.3  & 30.0  & 7.3  & 6.1  & 21.2  & 35.6  & 3.3  & 2.6  & 4.9  & 20.4 \\
Llama-3.3$_{w/ RAPM}$       & 70B & 3.8  & 2.7  & 11.3  & 24.3  & 29.6  & 25.0  & 21.7  & 32.4  & 18.5  & 15.5  & 19.7  & 33.1  & 9.1  & 7.3  & 8.2  & 23.7 \\
Gemini-3-Flash$_{w/ RAPM}$  & N/A  & 9.4  & 8.1  & 18.4  & 28.5  & 30.5  & 25.9  & 19.9  & 31.4  & 19.2  & 16.2  & 23.3  & 39.7  & 14.1  & 11.8  & 12.8  & 32.8 \\

\rowcolor{lightgray} \multicolumn{18}{l}{\textit{\textbf{Our method}}} \\
Llama-3.1$_{w/ 2D-ProteinRAG}$       & 8B & \textbf{44.3}  & \textbf{42.7}  & \textbf{41.5}  & \textbf{48.6}  & 28.3  & 23.9  & 23.0  & 32.9  & \textbf{72.7}  & \textbf{66.5}  & \textbf{44.5}  & \underline{59.8}  & 41.6  & 33.6  & 14.5  & 30.6 \\
Llama-3.3$_{w/ 2D-ProteinRAG}$       & 70B & 21.7  & 19.1  & 21.9  & 39.5  & \textbf{45.4}  & \textbf{38.8}  & \textbf{28.0}  & 37.6  & 69.5  & 63.6  & 38.6  & 52.2  & 45.4  & 36.6  & 8.1  & 19.2 \\
Gemini-3-Flash$_{w/ 2D-ProteinRAG}$  & N/A  & \underline{36.5}  & \underline{31.4}  & 31.5  & \underline{42.4}  & \underline{43.1}  & 32.9  & 15.9  & 28.4  & \underline{70.6}  & \underline{64.4}  & \underline{44.0}  & \textbf{59.9}  & \textbf{47.3}  & \textbf{39.4}  & 18.1  & 40.5 \\

\bottomrule
\end{tabular}}
\end{table*}

\textbf{The dual-dimensional filtering strategy is critical for distilling high-fidelity information from the inherent noise and redundancy of raw homology retrieval.}
A direct comparison between the w/ BLAST baseline (naive retrieval) and 2D-ProteinRAG reveals the magnitude of this denoising power. While raw BLAST retrieval provides the necessary biological information, the resulting context is often overwhelmed by exhaustive and conflicting metadata, which frequently confuses the generative model. For instance, on the Mol-Instructions Protein Function task, the Llama-3.1 baseline using raw retrieval yields an E-BLEU4 of only 8.7, which surges to 35.4 when our 2D strategy is applied—a nearly four-fold improvement. An even more dramatic escalation is observed in the Catalytic Activity task on the Prot-Inst-OOD dataset, where the score increases from 12.1 to 66.5. These results confirm that our Horizontal alignment and Vertical denoising modules effectively bridge the gap between noisy database outputs and specific user intents, providing the necessary precision to render retrieved evidence truly actionable for complex reasoning.

\subsection{More Strict OOD Experiences}

To go beyond standard OOD evaluations and verify model reliability under extreme generalization scenarios, we implement a strict homology constraint protocol. We add a test using CD-HIT~\cite{10.1093/bioinformatics/btl158} to filter the retrieval database, deliberately excluding any entries that share more than specific sequence identity thresholds (i.e. $>80\%$, $>60\%$ and $>40\%$) \footnote{According to \cite{Murzin1995-lw}, a <30\% identity falls within the ``Twilight Zone'', representing evolutionarily unrelated sequences. Furthermore, the average sequence identity of the Mol-Prot-Inst-OOD datasets between train and test sequences is 59.95\% with more similar description.} with the query protein in Prot-Inst-OOD dataset and Llama-3.1-8B. This setting simulates "twilight zone" conditions where only distant homologs are available, allowing us to rigorously assess whether our dual-dimensional filtering strategy remains robust even when high-confidence biological evidence is scarce.

\begin{figure}[h]
  \centering
  \includegraphics[width=0.9\linewidth]{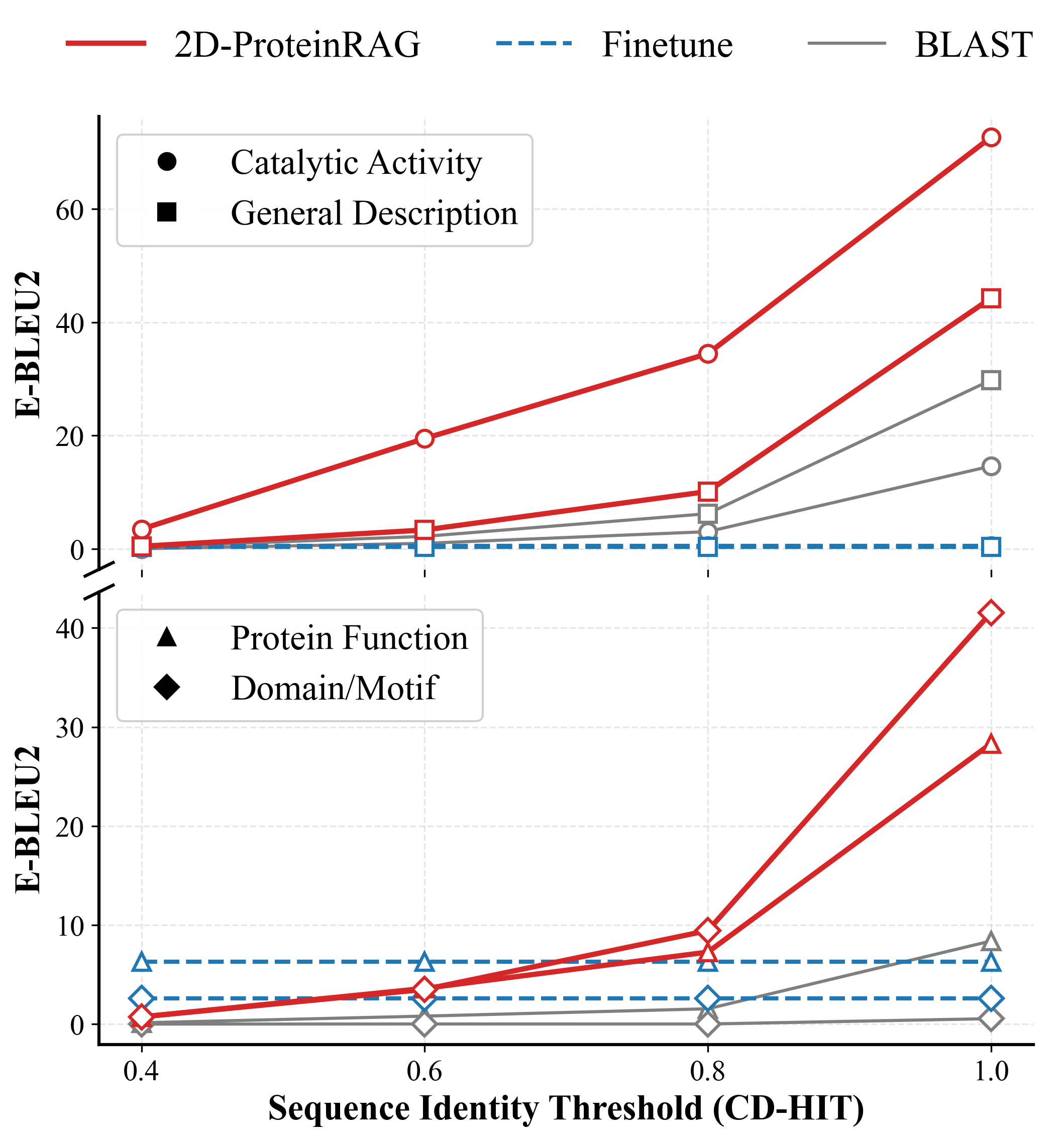}
  \caption{Robustness analysis under strict homology constraints (CD-HIT) across four tasks.}
  \label{cdhit_image}
\end{figure}

\textbf{2D-ProteinRAG maintains superior performance over naive retrieval and parametric knowledge across all homology constraints.} 
As illustrated in Figure \ref{cdhit_image}, while the performance of all methods inevitably declines as sequence identity decreases, our framework exhibits resilience. Specifically, as the threshold enters $\le 0.4$, the naive BLAST baseline suffers a catastrophic collapse to near-zero scores due to the overwhelming noise of distant homologs (0.02 in Catalytic Activity). In contrast, 2D-ProteinRAG effectively distills actionable context even from these remote relatives, maintaining functional signals where the baseline fails (3.48 in Catalytic Activity). Furthermore, our framework surpasses the fine-tuned baseline (Llama-3.1-8B, the dashed line) even under low-similarity constraints, achieving superior accuracy at identity thresholds as low as 40\% for tasks like Catalytic Activity. This result confirms that our framework can robustly synthesize functional insights even when high-confidence biological templates are unavailable.

\subsection{Further Experiences}

To further investigate the denoising efficacy of the individual components and the hyperparameter of the 2D-ProteinRAG framework, we conduct a series of detailed experiences. Specifically, we perform: (1) \textbf{Modular Ablation Studies} to quantify the contribution of each filtering dimension; (2) \textbf{Parameter Sensitivity Analysis} to evaluate the impact of retrieval context size $k$.

First, to isolate the contributions of our two-dimensional filtering strategy, we compare the full 2D-ProteinRAG framework against three variants: a naive \textit{w/ BLAST} baseline, a version with only Vertical Denoising (\textit{w/ denoising}), and a version with only Horizontal Alignment (\textit{w/ alignment}). The results on the Prot-Inst-OOD dataset are summarized in Table \ref{tab:modular_ablation}.

\begin{table}[htbp]
\centering
\small 
\caption{Ablation study on the Prot-Inst-OOD dataset quantifying the contribution of individual filtering components.}
\label{tab:modular_ablation}
\begin{tabular}{l cc cc}
\toprule
\multirow{2}{*}{\textbf{Model}}  & \multicolumn{2}{c}{\textbf{General Description}} & \multicolumn{2}{c}{\textbf{Protein Function}} \\
\cmidrule(lr){2-3} \cmidrule(lr){4-5}
 & \textbf{E-BL4} & \textbf{RG-L} & \textbf{E-BL4} & \textbf{RG-L} \\
\midrule

Llama-3.1$_{w/ BLAST}$              & 28.2  & 41.7  & 5.8  & 19.3 \\
Llama-3.1$_{w/ denoising}$  & 30.6  & 43.8  & 5.1  & 18.8 \\
Llama-3.1$_{w/ alignment}$ & 42.2  & 47.3  & 10.6  & 23.7 \\
Llama-3.1$_{w/ 2D-ProteinRAG}$       & 42.7  & 48.6  & 23.9  & 32.9 \\
\midrule

\multirow{2}{*}{\textbf{Model}} & \multicolumn{2}{c}{\textbf{Catalytic Activity}} & \multicolumn{2}{c}{\textbf{Domain/Motif}} \\
\cmidrule(lr){2-3} \cmidrule(lr){4-5}
& \textbf{E-BL4} & \textbf{RG-L} & \textbf{E-BL4} & \textbf{RG-L} \\
\midrule

Llama-3.1$_{w/ BLAST}$               & 12.1  & 21.0  & 0.3  & 8.1  \\
Llama-3.1$_{w/ denoising}$  & 13.2  & 22.5  & 0.4  & 8.1  \\
Llama-3.1$_{w/ alignment}$ &  65.4  & 58.4  & 30.3  & 28.6  \\
Llama-3.1$_{w/ 2D-ProteinRAG}$       &  66.5  & 59.8  & 33.6  & 30.6 \\

\bottomrule
\end{tabular}
\end{table}

\textbf{Both filtering dimensions are indispensable, as Horizontal Alignment ensures fine-grained intent-relevance while Vertical Denoising resolves adversarial functional contradictions.}
In Table \ref{tab:modular_ablation}, removing either dimension leads to significant performance degradation, validating their synergistic roles in the pipeline. The Horizontal Fine-grained Attribute Alignment serves as the primary engine for performance recovery. By selectively extracting task-relevant metadata, it provides the fine-grained information necessary for the LLM to follow specific instructions. This is most striking in the Catalytic Activity task, where adding the horizontal filter to the naive BLAST baseline triggers a massive surge in E-BLEU4 from 12.1 to 65.4, confirming its ability to find the key points of exhaustive database entries. Building upon this, the Vertical Homology-based Semantic Denoising further refines the context by resolving conflicting or adversarial signals across homologs. While horizontal filtering identifies the correct attributes, our vertical module addresses functional contradictions by enforcing a structural-anchored consensus. This effect is most evident in the Protein Function task. While horizontal alignment alone improves the score from 5.8 to 10.6, the addition of vertical denoising further propels it to 23.9, a 125\% relative increase. This demonstrates that resolving inter-homolog contradictions is critical for accurate reasoning, and only the full 2D-ProteinRAG framework ensures the retrieved evidence is both intent-aligned and biologically consistent.

\begin{figure}[h]
  \centering
  \includegraphics[width=0.9\linewidth]{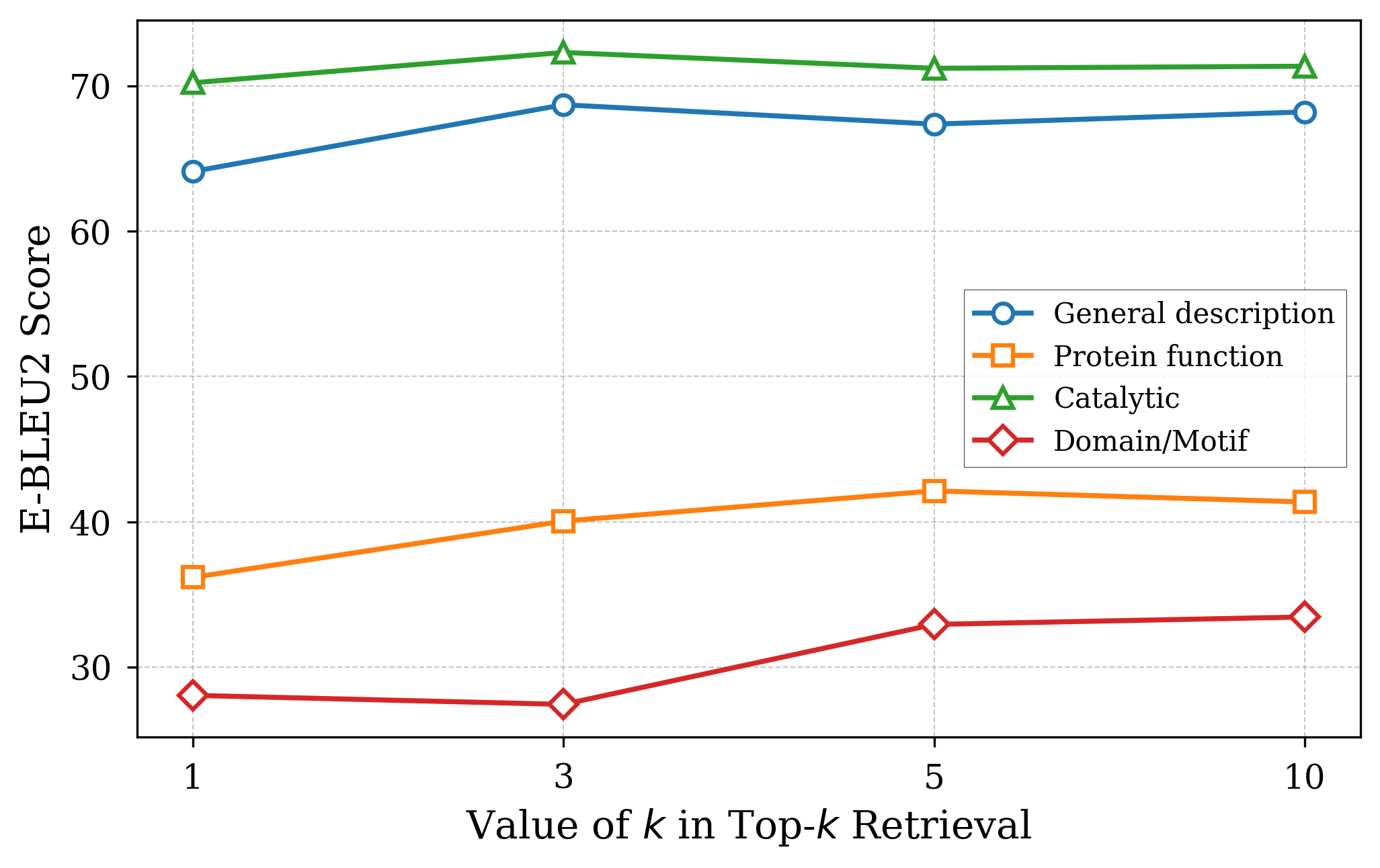}
  \setlength{\abovecaptionskip}{0pt}   
\setlength{\belowcaptionskip}{0pt}
  \caption{Impact of retrieval number $k$ on performance}
  \label{topk_ablation_image}
\end{figure}

Then, we investigate the impact of the retrieval context size $k$ on the overall performance of 2D-ProteinRAG. To determine the optimal balance between information richness and potential retrieval noise, we evaluate the framework across varying numbers of homologs, $k \in \{1, 3, 5, 10\}$. The resulting performance trends across all four downstream tasks are illustrated in Figure \ref{topk_ablation_image}.

\textbf{Top-$k$ sensitivity analysis reveals a task-dependent inform-ation-complexity trade-off, where challenging functional queries benefit from a broader evolutionary consensus.} 
In Figure \ref{topk_ablation_image}, the optimal retrieval size $k$ is intrinsically linked to the inherent complexity of the biological task. For cognitively demanding tasks such as Protein Function and Catalytic Activity, we observe that performance continues to scale up to $k=5$. This suggests that for complex functional inference, the model requires a more comprehensive evolutionary landscape to synthesize a robust consensus and resolve biological ambiguities. In contrast, for relatively straightforward tasks like General Description, performance reaches a saturation point at $k=3$, where additional homologs provide marginal gains but increase the risk of introducing irrelevant noise. Overall, while $k=5$ facilitates peak accuracy for high-complexity reasoning, $k=3$ emerges as a robust and computationally efficient heuristic across the entire benchmark. \footnote{More ablation study and case study can be found in Appendix \ref{app:exp}.}

\section{Conclusion}
In this paper, we presented 2D-ProteinRAG, a framework that bridges the gap between raw biological workflows and LLM reasoning through a dual-dimensional filtering strategy. By integrating the gold-standard BLAST tool directly into the RAG pipeline, our approach overcomes the OOD generalization bottlenecks inherent in closed-set training methods. Our strategy effectively extracts high quality biological information from noisy retrieval results by aligning metadata with specific instructions (Horizontal Alignment) and resolving functional contradictions across homologs via structure-anchored clustering (Vertical Denoising). Extensive evaluations on both ID and diverse biological OOD benchmarks demonstrate that 2D-ProteinRAG consistently achieves SOTA performance, showing that the synergy between open-world retrieval and expert-like denoising is essential for robust protein discovery. \footnote{The Limitations and Ethical Considerations are in Appendix \ref{app:lim}.}

\section*{GenAI Disclosure}

This work utilized Generative AI services to enhance the readability and flow of the text. The AI was prompted solely to rephrase sentences for academic clarity and correct typographical errors.

\bibliographystyle{ACM-Reference-Format}
\bibliography{sample-base}

\clearpage

\appendix

\section{Pseudocode}
\label{app:Pseudocode}
The pseudocode of our whole method can be found in Algorithm \ref{alg:2d_proteinrag}.

\begin{algorithm}[H]
\caption{2D-ProteinRAG Inference Pipeline}
\label{alg:2d_proteinrag}
\begin{algorithmic}[1]
\REQUIRE Protein sequence $x_p$, User instruction $x_i$
\REQUIRE Comprehensive Biological Database $\mathcal{D}$, Trained Horizontal Filter $\mathcal{M}_S$, Generative LLM $\mathcal{M}$

\STATE \textbf{Phase I: Raw Homology Retrieval}
\STATE Retrieve homologs: $\mathcal{H} \leftarrow \text{BLAST}(x_p; \mathcal{D})$
\STATE Initialize raw context: $\mathcal{C}_{raw} \leftarrow \emptyset$
\FOR{each homolog $h_j \in \mathcal{H}$}
    \STATE Retrieve annotation: $M_j \leftarrow \text{Lookup}(h_j; \mathcal{D})$
    \STATE Parse snippets: $C_j \leftarrow \{ (t_{j,m}, v_{j,m}) \}$ from $M_j$
    \STATE $\mathcal{C}_{raw} \leftarrow \mathcal{C}_{raw} \cup C_j$
\ENDFOR

\STATE \textbf{Phase II: Horizontal Filtering (Attribute Alignment)}
\STATE Initialize horizontal context: $\mathcal{C}_{hor} \leftarrow \emptyset$
\FOR{each snippet $(t, v) \in \mathcal{C}_{raw}$}
    \STATE Predict relevance score: $score \leftarrow \mathcal{M}_S(x_i, t)$
    \IF{$score > 0.5$}
        \STATE $\mathcal{C}_{hor} \leftarrow \mathcal{C}_{hor} \cup \{ (t, v) \}$
    \ENDIF
\ENDFOR

\STATE \textbf{Phase III: Vertical Filtering (Semantic Denoising)}
\STATE Extract descriptions: $\mathcal{V} \leftarrow \{ v \mid (t, v) \in \mathcal{C}_{hor} \}$
\STATE Compute embeddings: $E \leftarrow \text{Encoder}(\mathcal{V})$
\STATE Perform clustering: $\mathcal{K} = \{K_1, \dots, K_N\} \leftarrow \text{DBSCAN}(E)$
\STATE Identify top homolog: $h_{best} \leftarrow \mathcal{H}[0]$
\STATE Select Anchor Cluster:
\STATE \quad $K^* \leftarrow \{ K_n \in \mathcal{K} \mid \exists v \in K_n \text{ derived from } h_{best} \}$
\STATE Final refined context: $\hat{\mathcal{C}} \leftarrow \text{Text}(K^*)$

\STATE \textbf{Final Generation}
\STATE Generate response: $y \leftarrow \mathcal{M}(x_i, x_p, \hat{\mathcal{C}})$
\RETURN $y$
\end{algorithmic}
\end{algorithm}

\section{Dataset Details}
\label{app:data}
The sample statistics for the training and test sets of the two dataset categories utilized in this study are presented in Table \ref{tab:dataset_stats}.

\begin{table}[h]
\centering
\small
\caption{Statistics of the datasets used in experiments.}
\label{tab:dataset_stats}
\begin{tabular}{l c c c}
\toprule
\textbf{Dataset / Task} & \textbf{Train} & \textbf{Test} & \textbf{Identity} \\
\midrule
\multicolumn{3}{l}{\textbf{Mol-Instructions (Protein)}} & 71.31\% \\
\quad Protein Function & 110,689 & 3,494 \\
\quad Catalytic Activity & 51,573 & 1,601 \\
\quad Domain / Motif & 43,700 & 1,400 \\
\quad Functional Desc. & 83,939 & 2,633 \\
\midrule
\multicolumn{3}{l}{\textbf{OOD Datasets}} & 59.95\% \\
\quad Protein Function & 108,696 & 5,487 \\
\quad Domain / Motif & 42,368 & 2,732 \\
\quad Catalytic Activity & 51,187 & 1,987 \\
\quad General Function & 82,275 & 4,297 \\
\bottomrule
\end{tabular}
\end{table}

\section{Hyperparameters}
\label{app:hyp}
For all single-modal baseline models requiring fine-tuning, we adopted the identical hyperparameter configurations presented in Table~\ref{tab:baseline_hyperparams}, consistent with the settings reported in ~\cite{wu-etal-2025-rethinking-text}. For the multi-modal baseline \textit{Protein2Tex}\cite{jararweh-etal-2025-protein2text}, the training duration was set to 2 epochs, to align with the fine-tuning epochs of other baseline models, while all other parameters followed the settings described in its original paper. Furthermore, all RAG-based baseline models utilized the inference parameter settings detailed in Table~\ref{tab:inference_hyperparams}.

\begin{table}[htbp]
\centering
\small
\caption{Hyperparameter settings for fine-tuning baselines.}
\label{tab:baseline_hyperparams}
\begin{tabular}{l c}
\toprule
\textbf{Hyper-parameter} & \textbf{Value} \\
\midrule
\multicolumn{2}{l}{\textit{Training Dynamics}} \\
Learning Rate & 1e-4 \\
Batch Size (per device) & 2 \\
Gradient Accumulation Steps & 8 \\
Number of Epochs & 2 \\
Max Sequence Length & 2048 \\
Optimizer & AdamW\\
LR Scheduler Type & Cosine \\
Warmup Ratio & 0.1 \\
Weight Decay & 0.01 \\
Precision & BF16 \\
\midrule
\multicolumn{2}{l}{\textit{LoRA Configuration}} \\
LoRA Rank ($r$) & 8 \\
LoRA Alpha ($\alpha$) & 32 \\
LoRA Dropout & 0.05 \\
\midrule
\multicolumn{2}{l}{\textit{Hardware \& Infrastructure}} \\
DeepSpeed Config & ZeRO Stage-2 \\
Number of GPUs & 4 \\
Seed & 42 \\
\bottomrule
\end{tabular}
\end{table}

\begin{table}[htbp]
\centering
\small
\caption{Hyperparameter settings for baseline inference.}
\label{tab:inference_hyperparams}
\begin{tabular}{l c}
\toprule
\textbf{Hyperparameter} & \textbf{Value} \\
\midrule
Temperature & 0.7 \\
Top-$p$ (Nucleus Sampling) & 0.9 \\
Max Generation Tokens & 2048 \\
Presence Penalty & 0.0 \\
Frequency Penalty & 0.0 \\
\bottomrule
\end{tabular}
\end{table}

\section{Implementation Details}
\label{app:imp}
The temperature of final generation is 0.7 and a maximum token limit of 2048. All experiments were conducted on 4 NVIDIA H100 GPUs.

\subsection{Acquisition of BLAST Toolkit and Sequence Annotation Databases}

To ensure reproducibility, we detail the data acquisition and setup process as follows. First, we downloaded the BLAST+ executable binaries from the official NCBI website \footnote{https://blast.ncbi.nlm.nih.gov/Blast.cgi}. Next, we obtained the pre-formatted Swiss-Prot database from the same source. Specifically, we selected the manually annotated and reviewed UniProtKB/Swiss-Prot subset (Version 1.2, updated on September 2025).
We downloaded the corresponding textual annotation files from the official UniProt website \footnote{https://www.uniprot.org/} (version 45, updated on April 2025). To enable efficient retrieval based on Protein IDs, we constructed a local query index using the SeqIO.index\_db interface from Biopython. Finally, we downloaded and indexed the Gene Ontology (GO) database \cite{Harris2004-zl} to supplement the GO cross-references found in UniProt annotations, thereby maximizing the utilization of raw biological information.

It is worth noting that we exclusively utilized the Swiss-Prot database, distinguished by its manual review and high-quality annotations, rather than the TrEMBL dataset, which consists of unreviewed, computationally generated records. This selection ensures that our retrieval corpus contains the most accurate and rich functional descriptions available\cite{10.5555/3618408.3620023}.

\subsection{Details on Data Leakage Prevention during Retrieval}

To ensure a rigorous evaluation and prevent data leakage, we implemented a strict filtering protocol to exclude self-retrieval (i.e., retrieving the ground truth sequence itself from the database). Specifically, we define a retrieved sequence as identical to the input query if both its alignment length and identity length (count of identical matches) are exactly equal to the length of the input query sequence. During the testing phase, all such identical retrieval results were strictly excluded from the candidate pool to ensure that the model relies on homology-based reasoning rather than memorization.

\subsection{Details on Distillation Dataset Construction and Labeling}
\label{app:detail_m}
\textbf{Dataset Construction and Sampling.} We first retrieved the complete database entry for each sample sequence and parsed the raw text into structured Tag-Annotation key-value pairs. Then, to ensure a balanced training distribution, we cataloged the user instruction categories within each task. We then performed uniform sampling, collecting 100 data samples (consisting of Instruction and Key-Value sets) for each distinct instruction type. The data were divided into training and test sets in a 4:1 proportion.

\textbf{Information Gain (IG) Calculation.} To quantify the quality of each fine-grained annotation, we adopted the probability calculation method for LLM-generated sentences proposed in the \cite{wang-etal-2025-infogain}. This includes a sliding window smoothing mechanism and a token importance weighting algorithm. The specific steps are as follows:

Step 1: Sliding Window Smoothing: For each token $t_i$ in the generated answer sequence, its smoothed probability is calculated by averaging the probabilities of neighboring tokens within a window of size $W$:$$P_{\text{smooth}}(t_i) = \frac{1}{W} \sum_{j=i-\lfloor W/2 \rfloor}^{i+\lfloor W/2 \rfloor} p(t_j),$$where $p(t_j)$ represents the original token probability obtained by normalizing the LLM logits.

Step 2: Token Importance Weighting: It assigns higher importance weights to the first $k$ tokens when computing the final sequence probability $p_{\phi}(y|x)$. The calibrated confidence score is calculated as:$$p_{\phi}(y|x) = \prod_{i=1}^{k} (p_{\text{smooth}}(t_i))^{\omega_i \cdot \alpha} \cdot \prod_{j=k+1}^{|y|} (p_{\text{smooth}}(t_j))^{1-\alpha},$$where $\omega_i$ (set to 0.8) represents the importance weights for the first $k$ tokens, $\alpha$ is a weighting hyper-parameter, and $|y|$ is the total length of the answer.

Step 3: Finally, the algorithm computes the Information Gain (IG) to quantify the specific contribution of a retrieved annotation $d_i$ to the correct answer generation. This is defined as the difference between the model's confidence when conditioned on both the query and the document, versus the confidence conditioned on the query alone:$$\text{IG}(d_i|x) {=} p_{\phi}(y|x, d_i) - p_{\phi}(y|x),$$where $p_{\phi}(y|x, d_i)$ is the model's output confidence with the document, and $p_{\phi}(y|x)$ is the query-only confidence.

\textbf{Refinement for Multi-Attribute Tasks.} To address the partial relevance issue in multi-attribute tasks, we propose a Segment-wise Information Gain metric. Let the ground truth answer $y$ be decomposed into a sequence of $M$ semantic fragments $S_y = \{s_1, s_2, \dots, s_M\}$ (e.g., split by sentences). We define the effective information gain of a fine-grained annotation $d$ as the maximum gain it provides to any individual fragment $s_k$:$$\text{IG}_{\text{seg}}(d|x) = \max_{k \in \{1, \dots, M\}} \left( p_{\phi}(s_k|x, d) - p_{\phi}(s_k|x) \right),$$where $s_k$ denotes the $k$-th fragment of the ground truth $y$. $p_{\phi}(s_k|x, d)$ is the generation probability of fragment $s_k$ given the query $x$ and document $d$. $p_{\phi}(s_k|x)$ is the base generation probability of fragment $s_k$ given only the query. By maximizing over fragments, we ensure that an annotation $d$ is rewarded if it accurately aligns with any specific attribute described in $y$, regardless of its relevance to the remaining parts of the answer.

\textbf{Labeling Thresholds} According to the \cite{wang-etal-2025-infogain}, an $\text{IG} \gg 0$ indicates a significantly beneficial fine-grained document, $\text{IG} \approx 0$ suggests a noisy document, and $\text{IG} < 0$ implies a harmful document. For the distillation results using the Llama-3.2-1B model in this work, we set the selection threshold at 0.01. Annotations with an IG score greater than 0.01 were labeled as positive samples, while those below 0.01 were labeled as negative samples.

\section{More Experiments}
\label{app:exp}
We conducted additional ablation studies to evaluate the robustness of our method under different clustering configurations and teacher model choices for distillation.

\textbf{Effect of Top-$k$ Cluster Selection.} As shown in Table~\ref{tab:cluster_ablation}, we analyzed the impact of retrieving annotations from clusters corresponding to the Top-$k$ most similar sequences, which is an ablation study of change the $h_{best}$ in the Equation (\ref{eq:h}). The results indicate that selecting the Top-1 cluster yields the most stable denoising performance while maintaining high computational efficiency. Using more sequences to filter the results will further introduce noise to the results. Consequently, in our main experiments, we adopted the annotations from the single cluster containing the sequence with the highest similarity as the output for the vertical filter.

\textbf{Effect of Teacher Model Selection.} We further evaluated the downstream generation performance when distilling from different teacher models, as presented in Table~\ref{tab:teacher_model_ablation_mol} and Table \ref{tab:teacher_model_ablation_ood}. The model distilled from Llama-3.1-1B demonstrated superior performance combined with greater resource efficiency. This maybe come from that the 1B model is the most weak model in the Llama family, which will secure the lower bound of the need of the information.Therefore, it was selected as the backbone implementation for our main experiments.

\begin{table}[htbp]
\centering
\small 
\caption{Impact of Top K cluster selection on performance. Evaluated on Prot-Inst-OOD dataset.}
\label{tab:cluster_ablation}
\begin{tabular}{l cc cc}
\toprule
\multirow{2}{*}{\textbf{Model}}  & \multicolumn{2}{c}{\textbf{General Description}} & \multicolumn{2}{c}{\textbf{Protein Function}} \\
\cmidrule(lr){2-3} \cmidrule(lr){4-5}
 & \textbf{E-BL4} & \textbf{RG-L} & \textbf{E-BL4} & \textbf{RG-L} \\
\midrule

2D-ProteinRAG$_{Top1-cluster}$  & 41.4  & 48.4  & 23.7  & 32.6   \\
2D-ProteinRAG$_{Top2-cluster}$  & 43.3  & 47.0  & 23.4  & 32.1   \\
2D-ProteinRAG$_{Top3-cluster}$  & 41.2  & 45.5  & 22.9  & 31.5   \\
2D-ProteinRAG$_{Top4-cluster}$  & 39.9  & 45.0  & 22.8  & 31.2   \\
\midrule

\multirow{2}{*}{\textbf{Model}} & \multicolumn{2}{c}{\textbf{Catalytic Activity}} & \multicolumn{2}{c}{\textbf{Domain/Motif}} \\
\cmidrule(lr){2-3} \cmidrule(lr){4-5}
& \textbf{E-BL4} & \textbf{RG-L} & \textbf{E-BL4} & \textbf{RG-L} \\
\midrule

2D-ProteinRAG$_{Top1-cluster}$  & 64.9  & 56.8  & 32.4  & 28.8  \\
2D-ProteinRAG$_{Top2-cluster}$  & 59.4  & 53.8  & 34.2  & 27.5  \\
2D-ProteinRAG$_{Top3-cluster}$  & 58.5  & 52.5  & 35.6  & 27.1  \\
2D-ProteinRAG$_{Top4-cluster}$  & 59.6  & 52.5  & 35.5  & 27.1 \\

\bottomrule
\end{tabular}
\end{table}

We also show a case study of our framework in Figure \ref{fig:case_study}, which gives the path of how the framework gets the right answer step by step.
Initially, Top-K homologs are retrieved based on the input protein sequence. As shown in the red panel, this raw context contains extensive biological details (e.g., Function, Pathway, Subcellular Location). This information is excessively lengthy and noisy, containing unrelated data that could distract the model.
To address the noise, we apply Horizontal Filtering. Guided by the specific attribute emphasized in the user instruction (i.e., "Catalytic Activity"), this step employs fine-grained alignment to extract only the relevant text segments regarding catalytic activity from each homolog, effectively discarding irrelevant biological descriptions (Orange panel). 
Following attribute extraction, Vertical Filtering is applied (Blue panel). This step selects and consolidates the most representative evidence snippets (e.g., retaining key homologs like Q55C17 and Q9N5Y2) to construct a concise and highly relevant "Final Input."
Finally (Green panel), the model utilizes this dual-filtered context to generate a precise description of the enzyme's chemical reaction, which accurately matches the Ground Truth label.

\begin{table}[htbp]
\centering
\small 
\caption{Impact of different teacher models on distilling the Horizontal Fine-grained Attribute Alignment module (Mol-Instructions dataset).}
\label{tab:teacher_model_ablation_mol}
\begin{tabular}{l cc cc}
\toprule
\multirow{2}{*}{\textbf{Model}}  & \multicolumn{2}{c}{\textbf{General Description}} & \multicolumn{2}{c}{\textbf{Protein Function}} \\
\cmidrule(lr){2-3} \cmidrule(lr){4-5}
 & \textbf{E-BL4} & \textbf{RG-L} & \textbf{E-BL4} & \textbf{RG-L} \\
\midrule

Llama3.2-1B  & 67.3  & 62.9  & 35.4  & 36.3   \\
Llama3.2-3B  & 51.9  & 56.5  & 47.2  & 42.4   \\
Llama3.1-8B  & 51.7  & 56.5  & 43.0  & 39.9   \\

\midrule
\multirow{2}{*}{\textbf{Model}} & \multicolumn{2}{c}{\textbf{Catalytic Activity}} & \multicolumn{2}{c}{\textbf{Domain/Motif}} \\
\cmidrule(lr){2-3} \cmidrule(lr){4-5}
& \textbf{E-BL4} & \textbf{RG-L} & \textbf{E-BL4} & \textbf{RG-L} \\
\midrule

Llama3.2-1B  & 66.3  & 60.5  & 24.1  & 26.1   \\
Llama3.2-3B  & 40.1  & 36.9  & 36.3  & 29.3   \\
Llama3.1-8B  & 65.2  & 61.2  & 35.3  & 28.8   \\

\bottomrule
\end{tabular}
\end{table}

\begin{table}[htbp]
\centering
\small 
\caption{Impact of different teacher models on distilling the Horizontal Fine-grained Attribute Alignment module (Prot-Inst-OOD dataset).}
\label{tab:teacher_model_ablation_ood}
\begin{tabular}{l cc cc cc cc}
\toprule
\multirow{2}{*}{\textbf{Model}}  & \multicolumn{2}{c}{\textbf{General Description}} & \multicolumn{2}{c}{\textbf{Protein Function}} \\
\cmidrule(lr){2-3} \cmidrule(lr){4-5}
 & \textbf{E-BL4} & \textbf{RG-L} & \textbf{E-BL4} & \textbf{RG-L} \\
\midrule

Llama3.2-1B  & 42.7  & 48.6  & 23.9  & 32.9   \\
Llama3.2-3B  & 41.4  & 48.4  & 28.6  & 36.6   \\
Llama3.1-8B  & 30.2  & 43.9  & 33.7  & 38.3   \\

\midrule
\multirow{2}{*}{\textbf{Model}}  & \multicolumn{2}{c}{\textbf{Catalytic Activity}} & \multicolumn{2}{c}{\textbf{Domain/Motif}} \\
\cmidrule(lr){2-3} \cmidrule(lr){4-5}
 & \textbf{E-BL4} & \textbf{RG-L} & \textbf{E-BL4} & \textbf{RG-L} \\

Llama3.2-1B  & 66.5  & 59.8  & 33.6  & 30.6   \\
Llama3.2-3B  & 35.9  & 37.3  & 28.8  & 28.7   \\
Llama3.1-8B  & 58.0  & 52.2  & 32.5  & 30.1   \\

\bottomrule
\end{tabular}
\end{table}

\begin{figure*}[t]
\centering
\begin{tcolorbox}[
    colback=white, 
    colframe=gray!50, 
    boxrule=0.5mm, 
    arc=3mm,
    title=\textbf{Input-Output Example: 2D-ProteinRAG Inference Pipeline},
    coltitle=black,
    colbacktitle=gray!15,
    fonttitle=\large\bfseries
]

\begin{tcolorbox}[colback=mygray, colframe=gray, title=\small \textbf{USER INPUT}, sharp corners]
    \textbf{Instruction:} Examine the provided protein sequence and determine the catalytic activity of the enzyme it represents, focusing on the chemical reaction it promotes: \\
    \textbf{Sequence:} \texttt{MKVTVVS(omitted for brevity) ...PPFL}
\end{tcolorbox}

\begin{tcolorbox}[colback=myred, colframe=red!60!black, title=\small \textbf{PHASE I: RAW RETRIEVED CONTEXT (Top-K Homologs)}, sharp corners]
    \scriptsize
    \textbf{Homolog 1 (Q55C17):} [
  "FUNCTION: Catalyzes the last of the four reactions of the long-chain fatty acids elongation cycle. This endoplasmic reticulum-bound enzymatic process, allows the addition of 2 carbons to the chain of long- and very long-chain fatty acids/VLCFAs per cycle. This enzyme reduces the trans-2,3-enoyl-CoA fatty acid intermediate to an acyl-CoA that can be further elongated by entering a new cycle of elongation. Thereby, it participates in the production of VLCFAs of different chain lengths that are involved in multiple biological processes as precursors of membrane lipids and lipid mediators. CATALYTIC ACTIVITY: Reaction=a very-long-chain 2,3-saturated fatty acyl-CoA + NADP(+) = a   very-long-chain (2E)-enoyl-CoA + NADPH + H(+); PATHWAY: Lipid metabolism; fatty acid biosynthesis. SUBCELLULAR LOCATION: Endoplasmic reticulum membrane ; Multi-pass membrane protein SIMILARITY: Belongs to the steroid 5-alpha reductase family. Molecular Function: oxidoreductase activity; very-long-chain enoyl-CoA reductase activity Biological Process: fatty acid elongation; steroid biosynthetic process; very long-chain fatty acid biosynthetic process Cellular Component: endoplasmic reticulum membrane Domain/motif: Not found"
] \\
    \textbf{Homolog 2 (Q3ZCD7):} [
  "FUNCTION: Involved in both the production of very long-chain fatty acids for sphingolipid synthesis and the degradation of the sphingosine moiety in sphingolipids through the sphingosine 1-phosphate metabolic pathway (By similarity). Catalyzes the last of the four reactions of the long-chain fatty acids elongation cycle (By similarity). This endoplasmic reticulum-bound enzymatic process, allows the addition of 2 carbons to the chain of long- and very long-chain fatty acids/VLCFAs per cycle (By similarity). This enzyme reduces the trans-2,3-enoyl-CoA fatty acid intermediate to an acyl-CoA that can be further elongated by entering a new cycle of elongation (By similarity). Thereby, it participates in the production of VLCFAs of different chain lengths that are involved in multiple biological processes as precursors of membrane lipids and lipid mediators (By similarity). Catalyzes the saturation step of the sphingosine 1-phosphate metabolic pathway, the conversion of trans-2-hexadecenoyl-CoA to palmitoyl-CoA (By similarity). CATALYTIC ACTIVITY: Reaction=a very-long-chain 2,3-saturated fatty acyl-CoA + NADP(+) = a   very-long-chain (2E)-enoyl-CoA + NADPH + H(+);      PhysiologicalDirection=right-to-left; CATALYTIC ACTIVITY: Reaction=octadecanoyl-CoA + NADP(+) = (2E)-octadecenoyl-CoA + NADPH +   H(+);     PhysiologicalDirection=right-to-left; CATALYTIC ACTIVITY: Reaction=(2E,7Z,10Z,13Z,16Z)-docosapentaenoyl-CoA + NADPH + H(+) =   (7Z,10Z,13Z,16Z)-docosatetraenoyl-CoA + NADP(+);       PhysiologicalDirection=left-to-right; CATALYTIC ACTIVITY: Reaction=(2E,7Z,10Z,13Z,16Z,19Z)-docosahexaenoyl-CoA + NADPH + H(+) =   (7Z,10Z,13Z,16Z,19Z)-docosapentaenoyl-CoA + NADP(+);       PhysiologicalDirection=left-to-right; CATALYTIC ACTIVITY: Reaction=(2E,8Z,11Z,14Z)-eicosatetraenoyl-CoA + NADPH + H(+) =   (8Z,11Z,14Z)-eicosatrienoyl-CoA + NADP(+);     PhysiologicalDirection=left-to-right; CATALYTIC ACTIVITY: Reaction=(2E)-hexadecenoyl-CoA + NADPH + H(+) = hexadecanoyl-CoA +   NADP(+);     PhysiologicalDirection=left-to-right; PATHWAY: Lipid metabolism; fatty acid biosynthesis. PATHWAY: Lipid metabolism; sphingolipid metabolism. SUBUNIT: Interacts with ELOVL1 and LASS2. SUBCELLULAR LOCATION: Endoplasmic reticulum membrane ; Multi-pass membrane protein PTM: Glycosylated. SIMILARITY: Belongs to the steroid 5-alpha reductase family. Molecular Function: oxidoreductase activity; very-long-chain enoyl-CoA reductase activity Biological Process: fatty acid elongation; sphingolipid metabolic process; steroid biosynthetic process; very long-chain fatty acid biosynthetic process Cellular Component: endoplasmic reticulum; endoplasmic reticulum membrane Domain/motif: Not found"
] \\
    \textbf{Homolog 3:} ...
\end{tcolorbox}

\begin{tcolorbox}[colback=orange!10, colframe=orange!80!black, title=\small \textbf{PHASE II: HORIZONTAL FILTERED CONTEXT (Attribute Alignment)}, sharp corners]
    \scriptsize
    \textbf{Homolog 1 (Q55C17):} [
  "CATALYTIC ACTIVITY: Reaction=a very-long-chain 2,3-saturated fatty acyl-CoA + NADP(+) = a   very-long-chain (2E)-enoyl-CoA + NADPH + H(+);"
] \\
    \textbf{Homolog 2 (Q3ZCD7):} [
  "CATALYTIC ACTIVITY: Reaction=a very-long-chain 2,3-saturated fatty acyl-CoA + NADP(+) = a   very-long-chain (2E)-enoyl-CoA + NADPH + H(+);      PhysiologicalDirection=right-to-left; CATALYTIC ACTIVITY: Reaction=octadecanoyl-CoA + NADP(+) = (2E)-octadecenoyl-CoA + NADPH +   H(+);     PhysiologicalDirection=right-to-left; CATALYTIC ACTIVITY: Reaction=(2E,7Z,10Z,13Z,16Z)-docosapentaenoyl-CoA + NADPH + H(+) =   (7Z,10Z,13Z,16Z)-docosatetraenoyl-CoA + NADP(+);       PhysiologicalDirection=left-to-right; CATALYTIC ACTIVITY: Reaction=(2E,7Z,10Z,13Z,16Z,19Z)-docosahexaenoyl-CoA + NADPH + H(+) =   (7Z,10Z,13Z,16Z,19Z)-docosapentaenoyl-CoA + NADP(+);       PhysiologicalDirection=left-to-right; CATALYTIC ACTIVITY: Reaction=(2E,8Z,11Z,14Z)-eicosatetraenoyl-CoA + NADPH + H(+) =   (8Z,11Z,14Z)-eicosatrienoyl-CoA + NADP(+);     PhysiologicalDirection=left-to-right; CATALYTIC ACTIVITY: Reaction=(2E)-hexadecenoyl-CoA + NADPH + H(+) = hexadecanoyl-CoA +   NADP(+);     PhysiologicalDirection=left-to-right;"
] \\
    \textbf{Homolog 3:} ...
\end{tcolorbox}

\begin{tcolorbox}[colback=myblue, colframe=blue!60!black, title=\small \textbf{PHASE III: VERTICAL FILTERED CONTEXT (Final Input)}, sharp corners]
    \small
    \textbf{Homolog 1 (Q55C17):} [
  "CATALYTIC ACTIVITY: Reaction=a very-long-chain 2,3-saturated fatty acyl-CoA + NADP(+) = a   very-long-chain (2E)-enoyl-CoA + NADPH + H(+);"
] \\
    \textbf{Homolog 2 (Q9N5Y2):} [
      "CATALYTIC ACTIVITY: Reaction=a very-long-chain 2,3-saturated fatty acyl-CoA + NADP(+) = a   very-long-chain (2E)-enoyl-CoA + NADPH + H(+);"
    ] 
\\
\end{tcolorbox}

\begin{tcolorbox}[colback=mygreen, colframe=green!50!black, title=\small \textbf{GENERATION \& EVALUATION}, sharp corners]
    \textbf{Model Output:} An analysis of the protein sequence reveals that the enzyme's catalytic function corresponds to the chemical reaction: a very-long-chain 2,3-saturated fatty acyl-CoA + NADP(+) = a very-long-chain (2E)-enoyl-CoA + NADPH + H(+).\\
    \rule{\linewidth}{0.2pt} 
    \textbf{Label (Ground Truth):} Upon reviewing the provided protein sequence, the corresponding enzyme's catalytic activity is identified as the following chemical reaction: a very-long-chain 2,3-saturated fatty acyl-CoA + NADP(+) = a   very-long-chain (2E)-enoyl-CoA + H(+) + NADPH.
\end{tcolorbox}

\end{tcolorbox}
\caption{An illustrative example of the 2D-ProteinRAG inference process. The pipeline progressively filters noise from raw retrieval to generate accurate descriptions.}
\label{fig:case_study}
\end{figure*}

\section{Limitations and Ethical Considerations}
\label{app:lim}
\textbf{Limitations.} While our proposed framework, 2D-ProteinRAG, demonstrates state-of-the-art performance, we acknowledge several limitations: Inference Latency: Compared to purely parametric models, the retrieval-augmented generation (RAG) process introduces additional computational overhead during inference, particularly when searching against large-scale protein databases. Hallucination Risks: Although RAG significantly mitigates hallucinations by grounding generation in retrieved context, there remains a residual risk of generating plausible but factually incorrect chemical reaction descriptions, which requires expert verification for critical downstream applications.

\textbf{Ethical Use of Data and Human Subjects.} We strictly adhere to the ACM Publications Policy on Research Involving Human Participants and Subjects. The datasets utilized in this study (e.g., UniProtKB, Swiss-Prot) are publicly available scientific repositories containing anonymized protein sequences and functional annotations. No human participants, clinical trials, or personally identifiable information (PII) were involved in the data collection or experimental processes of this research. All data usage complies with the licensing terms of the respective source databases.

\end{document}